# Smart Cities and Digital Twins in Lower Austria

Gabriela Viale Pereira*
Lucy Temple*
Thomas J. Lampoltshammer
gabriela.viale-pereira@donau-uni.ac.at
lucy.temple@donau-uni.ac.at
thomas.lampoltshammer@donau-uni.ac.at
University for Continuing Education Krems
Krems an der Donau, Austria

Lukas Daniel Klausner*
Thomas Delissen
Torsten Priebe
mail@l17r.eu
thomas.delissen@fhstp.ac.at
torsten.priebe@fhstp.ac.at
St. Pölten University of Applied Sciences
St. Pölten, Austria

## ABSTRACT

Smart city solutions require innovative governance approaches together with the smart use of technology, such as digital twins, by city managers and policymakers to manage the big societal challenges. The project Smart Cities aNd Digital Twins in Lower Austria (SCiNDTiLA) extends the state of the art of research in several contributing disciplines and uses the foundations of complexity theory and computational social science methods to develop a digital-twin-based smart city model. The project will also apply a novel transdisciplinary process to conceptualise sustainable smart cities and validate the smart city generic model. The outcomes will be translated into a roadmap highlighting methodologies, guidelines and policy recommendations for tackling societal challenges in smart cities with a focus on rescaling the entire framework to be transferred to regions, smaller towns and non-urban environments, such as rural areas and smart villages, in ways that fit the respective local governance, ethical and operational capacity context.

## 1 INTRODUCTION

This poster gives a brief introduction to the project **S**mart **Ci**ties a**N**d **D**igital **T**wins **i**n **L**ower **A**ustria (SCiNDTiLA, 2023–2026), in which we explore the concept of smart cities and how it can be transferred to smaller-scale urban and non-urban contexts by using digital twins technology and algorithmic decision support to assist policymakers in developing smart sustainable solutions.

SCiNDTiLA analyses the current state of research on smart cities and digital twins and adapts these technologies to the Lower Austrian context. Small Austrian cities and regions will be modelled as smart cities/regions using complexity-theoretical and computational social science methods, which can then serve as systems of socio-technical interaction to support policy decisions for sustainable local governance.

The objectives of the project are to (1) define the state of the art in the field of smart cities and identify those characteristics that can be transferred to smaller-scale urban and non-urban contexts; (2) develop a conceptual framework of sustainable local governance via digital twins to validate the smart city generic model through an innovative transdisciplinary process; (3) develop a digital-twin-based sustainable smart city and define different scenarios concerning challenges of good governance in smaller-scale urban and non-urban contexts; (4) implement the proof of concept use cases in Lower Austria and propose a roadmap highlighting methodologies, guidelines and policy recommendations on how smart and sustainable solutions in cities and regions shape inhabitants' perceptions of local governance.

The four objectives of the project altogether provide a comprehensive response to the main research question raised: How can the existing knowledge on smart cities be transferred to smaller-scale urban and non-urban contexts and how can technologies such as digital twins be used to support policymakers in developing smart sustainable solutions for cities and regions in Lower Austria? The sub-questions of the project are twofold: 1) In what way do existing standards, research, use cases and best practices of smart city initiatives need to be adapted to be useful for and applicable to smaller towns and regions? 2) How can a digital-twin-based smart city model support the formulation and integration of governance and sustainability with policy-making, and thus develop smart sustainable solutions for smaller-scale contexts?

The pursuit of our project's objectives requires the formalisation of the characteristics of a smart city and the operations that permit its simulation and the definition of methods for selecting policies among the many possibilities. This may help tackle so-called "wicked problems" such as those related to sustainability for smart cities. The roadmap will guide policy decision-making in applying the proposed method for assessing policies and their effectiveness according to the (social) changes in the state of the system (based on inhabitants' perceptions and by facilitating participation), i. e. the impact of the smart-city-related policies for sustainable development and fostering social cohesion.

Finally, we will rescale our policy recommendations to reach communities of practice across Lower Austria. Specifically, we consider regional policymakers and smart city managers as proxies for promoting collaborative decision-making and engaging inhabitants towards sustainable solutions.

## 2 METHODOLOGICAL APPROACH

Our theoretical background follows an interdisciplinary approach, combining social and technical sciences to address the new challenges of innovation, development and sustainability in smart cities. The problem that we address has been previously treated by the scientific literature according to three primary directions; (1) the study of governance and sustainability within smart cities [4, 7], (2) the use of digital twins or avatars for promoting inclusion and participation [6, 8], and (3) the use of scientific knowledge concerning the relationship between the changes in the state of a social system and its sustainability for the purpose of selecting policies [2].

The research method of this project follows a design science research approach [3] to define a smart city as a system (an IT artefact) to be evaluated in a given organisational context – in this case, the state of Lower Austria, or more precisely, with regard to the social impact upon Lower Austrian towns and regions [1]. The project will also apply a novel transdisciplinary process [5] to conceptualise sustainable smart cities and validate the smart city generic model, including the contextualisation and identification of the main societal challenges to be addressed as use cases in small cities and regions in Lower Austria. The output of this process (science–practice intersection) is an ontology on the representation of an abstract smart city as a generalised process that can be replicated.

To carry out this methodology, the project will carry out five tasks. **Task 1** focuses on conducting a systematic literature review and evidence assessment to establish a baseline for sustainable smart cities and regions. **Task 2** develops a conceptual framework for smart sustainable cities and engages with local stakeholders to identify urgent societal challenges and use cases. **Task 3** involves building a digital-twin-based smart city simulation model and validating the use cases based on the requirements elicitation for the Lower Austrian context. **Task 4** consists in conducting simulations to analyse stakeholders' perceptions, developing a proof of concept and drafting methodologies and guidelines for tackling the identified societal challenges in smart cities/regions. The final **Task 5** focuses on dissemination and sustainability: disseminating the project and creating durable links to public authorities to foster collaboration, establish a network for cooperation, and encourage long-term collaboration between city administrations and universities. Overall, the project aims to develop a replicable process for sustainable smart cities that can be applied to smaller-scale urban and non-urban contexts.

## 3 EXPECTED RESULTS

The outcomes of the project will take particular care to establish the link between sustainable smart cities/regions and their impacts on the public's perception of local governance and the way government tackles societal challenges. As the two agendas have thus far been developed in relative isolation, we believe that such an approach can make a unique contribution to improve the effectiveness of smart city solutions implemented in Lower Austria and beyond.

Our approach combines a systemic one, identifying interdependencies, critical components and stakeholders, and an operational upscaling to provide a roadmap for smart cities with application to both smaller-scale urban and non-urban environments. The roadmap will guide policy decision-making in applying the proposed method for assessing policies and their effectiveness according to the (social) changes in the state of the system (based on inhabitants' perceptions and facilitating participation), i. e. the impact of smart-city-related policies for sustainable development and fostering social cohesion.

## ACKNOWLEDGMENTS

This research was funded by the Gesellschaft für Forschungsförderung Niederösterreich (GFF NÖ) project GLF21-2-010 "Smart Cities and Digital Twins in Lower Austria". The financial support by the Gesellschaft für Forschungsförderung Niederösterreich is gratefully acknowledged.